\documentclass[twocolumn,10pt]{tsfp}
\usepackage{flushend}
\usepackage{graphicx}
\usepackage[authoryear,round]{natbib}
\usepackage{fancyhdr}
\usepackage{epstopdf}

\newcommand\Rey{\mbox{\textit{Re}}}  % Reynolds number
\newcommand\Fro{\mbox{\textit{Fr}}}  % Froude number, cf TeX's \Fr product
\newcommand{\appropto}{\mathrel{\vcenter{
			\offinterlineskip\halign{\hfil$##$\cr
				\propto\cr\noalign{\kern2pt}\sim\cr\noalign{\kern-2pt}}}}}

\graphicspath{ {./figures/}}

\title{Stratified wakes of a prolate spheroid at moderate angle of attack}

\author{Sheel Nidhan
	\affiliation{Mechanical and Aerospace Engineering\\
		University of California San Diego\\
		CA 92093\\
		snidhan@ucsd.edu
	}	
}

\author{Sanidhya Jain
	\affiliation{Mechanical and Aerospace Engineering\\
		University of California San Diego\\
		CA 92093\\
		saj004@ucsd.edu
	}	
}

\author{Jose L. Ortiz-Tarin
	\affiliation{Mechanical and Aerospace Engineering\\
		University of California San Diego\\
		CA 92093\\
		jlortiztarin@gmail.com
	}
}

\vspace{-5mm}
\author{Sutanu Sarkar
	\affiliation{Mechanical and Aerospace Engineering\\
		University of California San Diego\\
		CA 92093\\
		ssarkar@ucsd.edu
	}
}

\begin{document}

\maketitle   %Print title matter
\thispagestyle{fancy}

% Set the font to 9pt.
\fontsize{9}{11}\selectfont

%%%%%%%%%%%%%%%%%%%%%%%%%%%%%%%%%%%%%%%%%%%%%%%%%%%%%%%%%%%%%%%%%%%%%%
\section*{ABSTRACT}

Ocean submersibles and aerial vehicles often encounter a density-stratified environment whose effect on flow features is of interest. A 6:1  prolate spheroid of diameter $D$ with  velocity $U$ and at  a moderate angle of attack (AOA) of $10^\circ$ is taken as a canonical example of a submersible. Buoyancy effects are examined in a parametric LES study of the spheroid wake at $\Rey = UD/\nu = 5000$ where stratification is changed to cover a wide range of values of body Froude number  ($\Fro = U/ND$). The Froude number  measures buoyancy time scale (1/$N$ where $N$ is the buoyancy frequency)  relative to flow time scale ($D/U$).  The simulated cases range from a baseline case without stratification, i.e. $\Fro = \infty$,   to the substantial stratification level of  $\Fro = 1$.  The very near wake, just two body diameters aft of the trailing edge, is found to be substantially altered  at even the relatively weak stratification of $\Fro = 6$. Specifically,  the coherence of the streamwise vortex pair shed from the body is weakened,  the downward trajectory of the wake center is suppressed, and the mean/turbulence structure changes in the near wake.   Diagnosis of the vorticity transport equation reveals that the baroclinic torque becomes an important contributor to the balance of mean streamwise vorticity in the very near wake at $\Fro = 6$.   With increasing stratification, the wake topology changes significantly, e.g. the $\Fro =1$ case exhibits a secondary wake above the primary wake.
%
%This document is a template for preparing a manuscript for TSFP on A4 size paper.
%
%The maximum length of the manuscript is six (6) pages.
%
%Begin your abstract 115 mm from the top of the page as shown on this page. The ABSTRACT heading should be boldface in all capitals; it should be in a 9 pt. sans serif typeface (Helvetica) as shown above. The ABSTRACT section should not exceed 250 words. It should be flush left with the left margin. The abstract text should be in 9 pt. New Times Roman font, the same as the main text. The spacing to the next heading should be two (2) line spaces.

%%%%%%%%%%%%%%%%%%%%%%%%%%%%%%%%%%%%%%%%%%%%%%%%%%%%%%%%%%%%%%%%%%%%%%
%%%%%%%%%%%%%%%%%%%%%%%%%%%%%%%%%%%%%%%%%%%%%%%%%%%%%%%%%%%%%%%%%%%%%%
\section*{INTRODUCTION}
Despite its relevance to applications, the wake of a slender body  has not received much attention compared to that of a bluff body.  Aspects such as the length of the body and its wake, peculiarities of flow separation on a mildly curved body and the disparate scales between the boundary layer on the body and its far wake present challenges. For a body with uniform speed $U$ with respect to the ambient, the governing parameters in a homogeneous fluid are as follows: the angle of attack (AOA or $\alpha$), aspect ratio ($L/D$ which is larger than $O(1)$) and Reynolds number ($\Rey= UD/\nu$ or $\Rey_L = UL/\nu$). In a density-stratified background with buoyancy frequency $N$ defined by $N^2 = -(g/\rho_0) \partial \rho_b/\partial z$, the body Froude number ($\Fro = U/ND$) is an important determinant of wake properties.

The first experimental study of flow past a slender body at zero AOA dates back to \cite{Chevray1968} who investigated the wake of a 6:1 prolate spheroid at $\Rey = 4.5 \times 10^{5}$. This work and more recent studies of  axisymmetric slender bodies~\citep{Jimenez2010,Posa2016,Kumar2018,Ortiz2021} at $\Rey = O(10^5)$ have studied the flow into the near wake, $x/D < 20$.  The first study \citep{Ortiz2021} that probed slender-body far wake statistics was a LES that extended to $x/D = 80$. These simulations at $\Rey_L = 6 \times 10^5$ were performed with a trip to facilitate transition to turbulence in the boundary layer. 

Underwater bodies operate in an  environment whose density is stratified owing to temperature or salinity. Bluff-body wakes, particularly the sphere wake, have received much attention, first through laboratory studies  and then through simulations. Slender-body wakes in a stratified fluid have received scant attention, except for \cite{Meunier2004} who compared the evolution in the very far wake among  several body shapes that also included a 6:1 prolate spheroid (data shown after $x = 100D$).
The scale-resolving study \citep{Ortiz-tarin_stratified_2019} was the first simulation to account for background stratification  in slender-body flows. Their  LES was conducted for a 4:1 spheroid at zero angle of attack, $\Rey = 10^{4}$, and $\Fro = \infty, 3, 1$ and $0.5$. They analyzed the laminar BL evolution, force distribution, and the near- and far-field characteristics of the steady lee waves. In a followup study \citep{Ortiz-tarin_stratified_2022} of a more slender body - a 6:1 spheroid -  at  a higher $\Rey_L = 6 \times 10^5$ with turbulent boundary layer separation, several differences were found between the spheroid wake and the bluff-body wake of a disk at similar Reynolds and Froude number. The streamwise locations of the transition points in the multi-stage wake evolution~\citep{Spedding1997} were found to be very different than those for a bluff body.

\begin{figure}
\centering
\includegraphics[width=0.5\textwidth,keepaspectratio]{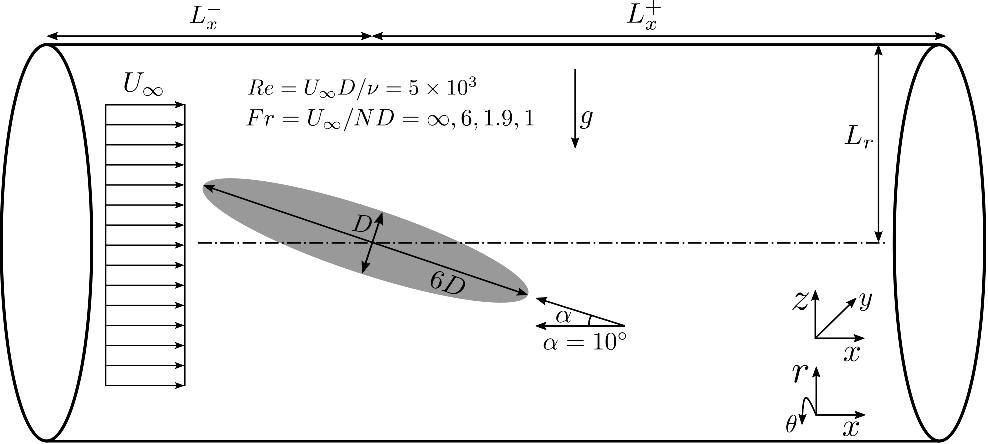}
\caption{Schematic of a spheroid at an angle of attack in the cylindrical computational domain.  $L_x^{-}$, $L_x^{+}$ and $L_r$ refer to the upstream, downstream and radial domain distance, respectively. $\Rey$, $\Fro$ and $\alpha$ correspond to the diameter-based Reynolds number, diameter-based Froude number and angle of attack, respectively.}         
\label{fig:setup}
\end{figure}

As noted above, slender-body wakes have been shown recently to exhibit buoyancy effects that are substantially different than bluff-body wakes when the flow is straight on. Little is known about slender body wakes at an angle of attack in a stratified enviroment. We are thus motivated to examine the wake of a slender body - a prolate 6:1 spheroid - at  an angle of attack (see schematic, Figure \ref{fig:setup}).  $\Rey = 5000$ is smaller than in our previous work so that the resolution is high, close to but not quite a DNS. 

\section*{NUMERICAL METHODOLOGY}

The numerical solver used for these simulations have been extensively validated in the past for body-inclusive simulations of stratified wakes (\cite{Ortiz-tarin_stratified_2019,chongsiripinyo_decay_2020}). The three-dimensional Navier-Stokes equations with the Boussinesq approximation in cylindrical coordinates is numerically solved.  A third-order Runge-Kutta method combined with  second-order Crank-Nicolson is employed  to advance the equations in time. Second-order-accurate central differences are used for the spatial derivatives in a staggered grid. The dynamic Smagorinsky model is used to account for subgrid fluctuations. The boundary conditions are Dirichlet at the inflow, convective outflow and Neumann at the radial boundary. A  sponge layer is added to the boundaries to avoid the spurious reflection of gravity waves.  An immersed boundary method is used to represent the body.  A synopsis of the flow solver in cylindrical coordinates, boundary conditions and immersed boundary method (IBM) approach is available in  \cite{chongsiripinyo_decay_2020}. 

The present work reports results at  $\Rey = 5 \times 10^{3}$ and  four stratification levels -  $\Fro = \infty, 6,1.9,$ and $1$ - at an angle of incidence $\alpha = 10^{\circ}$. For the stratified cases at $\alpha = 10^{\circ}$, radial and streamwise domains span  $0 \leq r/D \leq 53$ and $-30 \leq x/D \leq 50$, respectively. A large radial extent together with a sponge layer on the boundaries weaken the internal waves before they hit the end of the computational domain and hence control the amplitude of spurious reflected waves. The grid point distribution is as follows: $N_{r} = 1000$ in the  radial direction, $N_{\theta} = 128$ in the azimuthal direction, and $N_{x} = 3584$ in the streamwise direction. For the unstratified wake at $\alpha = 10^{\circ}$, radial and streamwise domains span $0 \leq r/D \leq 21$ and $-11 \leq x/D \leq 47$ while $N_r = 718, N_{\theta} = 256$, and $N_x = 2560$.

The boundary layer is laminar and is very well resolved with about 40 points across it. To assess grid quality in the wake, the ratio of grid spacing to the Kolmogorov length, denoted as $\eta = (\nu^3/\varepsilon)^{1/4}$, is computed  in all three directions. The ratios of streamwise ($\Delta x$), radial ($\Delta r$), and azimuthal ($r \Delta \theta$) grid spacing to $\eta$ stay  below $6$, $1$ and $5$, respectively, for all wakes, establishing that the current large eddy simulations are at high resolution.

\section*{Results}

The results  are  presented using a fixed orthogonal coordinate system  that is aligned with the uniform freestream in  the horizontal $x$ direction.  The origin is placed at the center of the 6:1 spheroid so that, at zero AOA, the trailing edge is at $x/D = 3 \cos (10^\circ) = 2.95$.    For statistics, a time averaging window of approximately $100 D/U$ is used. 
%In the final paper, values of  $\Rey $ up to 20,000 will be discussed to ascertain robustness of some of the new findings.

\begin{figure*}[ht]
\centering
\includegraphics[width=0.8\linewidth,keepaspectratio]{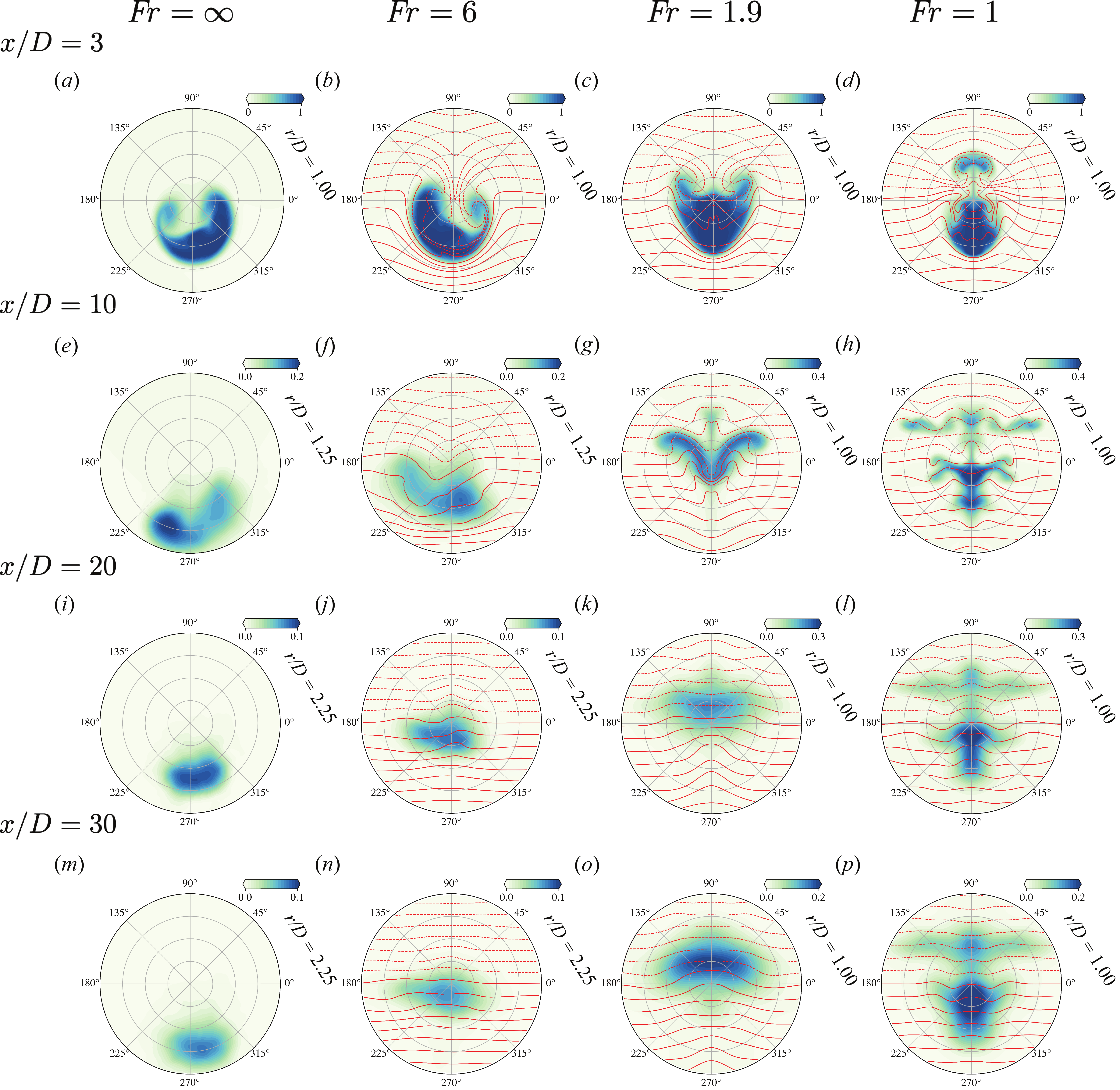}
\caption{Mean defect velocity ($U_d$) contours and isopycnals at $x/D = 3, 10, 20$ and $40$ (row-wise) for $\Fro = \infty, 6, 1.9$ and $1$ (column-wise). The radial domain is shown adjacent to each contour plot. For the stratified cases, body-generated steady lee waves in the domain are not visible due to our choice of contour ranges that allow us to focus specifically on the wake defect.} 
\label{fig:mean_defect_contours_aoa10}
\end{figure*}

\begin{figure*}[ht]
	\centering
	\includegraphics[width= 0.9 \linewidth,keepaspectratio]{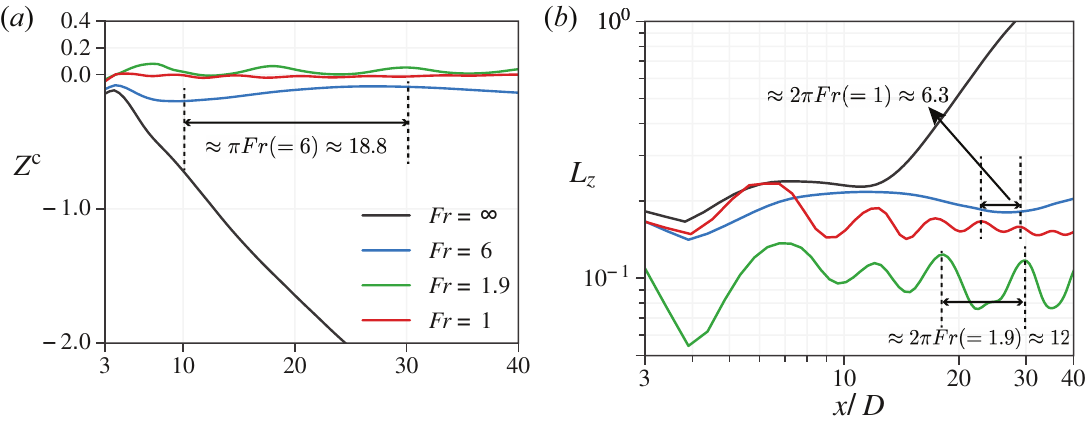}
	\caption{Evolution of overall wake geometry: (\textit{a}) the vertical location of wake center($Z^{c}$), and (\textit{b}) vertical ($L_z$) lengthscale.}  
	\label{fig:mean_length_mke_aoa10}
\end{figure*}

Flow separation (laminar at $\Rey = 5000$) from the inclined spheroid leads to a counter-rotating vortex pair at the body and in the near wake.The wake is turbulent.  Buoyancy modifies the vortex structures, the wake  and turbulence levels/anisotropy (not shown).

Figure \ref{fig:mean_defect_contours_aoa10} presents mean defect velocity ($U_d$) contours at various streamwise locations. In the stratified scenarios, mean isopycnals are superimposed on the $U_d$ contours. 
In the unstratified wake ($\Fro = \infty$), two distinct lobes are evident  (Figure \ref{fig:mean_defect_contours_aoa10}${a}$) at the trailing edge.  The asymmetry in the $U_d$ profile between the left and right lobes arises from asymmetric flow separation over the body. Asymmetric flow separation also occurs for $\Fro = 6$, but with the direction of asymmetry reversed. There is no preferred direction for lateral asymmetry in the flow configuration by design. The flow locks in to one of  two distinct reflectional-symmetry-breaking states, with each state being equally probable. The asymmetric lobes of  the $\Fro = \infty$ wake persist until at least $x/D = 10$ (figure \ref{fig:mean_defect_contours_aoa10}${e}$). However, by $x/D = 20$, these near-wake lobes disappear.  Another noteworthy observation is the continuous drift of the entire wake vertically downward.

The lateral asymmetry that is found here in the wake is worth discussion. 
Previous studies on flow past slender axisymmetric bodies at both high \citep{jiang_transitional_2015} and low-to-moderate \citep{ashok_asymmetries_2015} angles of attack have observed the emergence of lateral asymmetry in the wake. A three-dimensional linear stability analysis ~\citep{tezuka_three-dimensional_2006} of the flow past a 4:1 spheroid provided a theoretical basis for the observed non-intuitive lateral asymmetry of flow separation.  As $\Rey$ was increased in the $O(10^4)$ range,  the authors found transition from a symmetric steady to an asymmetric steady configuration at $\Rey = \Rey_{c1}$ and to an asymmetric oscillatory configuration at a larger $\Rey = \Rey_{c2}$.  For a 4:1 spheroid at AOA = 10$^\circ$, theory shows transition to an asymmetric steady wake at $\Rey_{c1} \approx 4000$ and, notably, the laboratory experiment  by the same authors show an asymmetric pattern of surface streamlines at $\Rey = 5500$ along with a sideways mean force.

The mean wake topology for the $\Fro = 6$ case is shown in figure \ref{fig:mean_defect_contours_aoa10} (second column). Although the $\Fro=6$ wake is weakly stratified,  differences  with the unstratified case occur quite early. While the unstratified wake continues to drift downward in the negative $z$ direction, the $\Fro =6$ wake remains closer to the centerline due to the inhibiting effects of buoyancy -- compare figure \ref{fig:mean_defect_contours_aoa10}($i,m$) with figure \ref{fig:mean_defect_contours_aoa10}($j,n$). 
%Plots of the streamwise vorticity (not shown) reveal that in the near wake at $x/D =10$,  while the streamwise vortex pair shed from the body remains coherent at $\Fro = \infty$,  it disintegrates at $Fr = 6$.  
 The isopycnals (constant-density lines) are strongly distorted by the wake at $x/D = 3$. As the wake progresses downstream, the isopycnals tend to return to their neutral position in the $\Fro = 6$ wake. 

In strongly stratified cases of $\Fro = 1.9$ and 1, the mean flow leaving the body is no longer laterally asymmetric, as seen in figure \ref{fig:mean_defect_contours_aoa10}($c,d$). The $\Fro = 1.9$ wake and its corresponding isopycnals exhibit strong lee-wave-induced oscillations as it progresses downstream. The $\Fro = 1$ wake is qualitatively distinct for the other cases in that  a secondary wake above the primary wake emerges near the body and persists downstream as shown in the rightmost column.

\begin{figure*}[ht!]
	\centering
	\includegraphics[width=\linewidth,keepaspectratio]{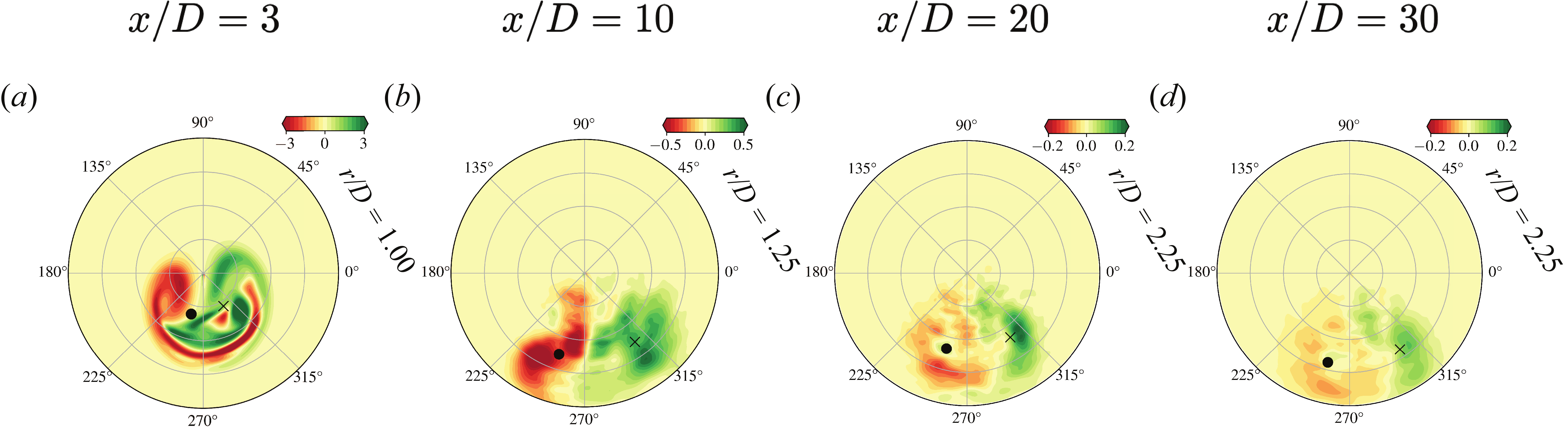}
	\caption{Mean streamwise vorticity $\langle \omega_x \rangle$ contours at $x/D = 3, 10, 20$ and $30$ for $\Fro = \infty$. $r/D$ denotes the radial extent of the contour at the respective $x/D$ locations. The black dot and cross represent the center of the negative and the positive vortex filament, respectively.}
	\label{fig:mean_omgx_contours_frinf}
\end{figure*}
\begin{figure*}
%\begin{figure*}[ht!]
	\centering
	\includegraphics[width=0.9\linewidth,keepaspectratio]{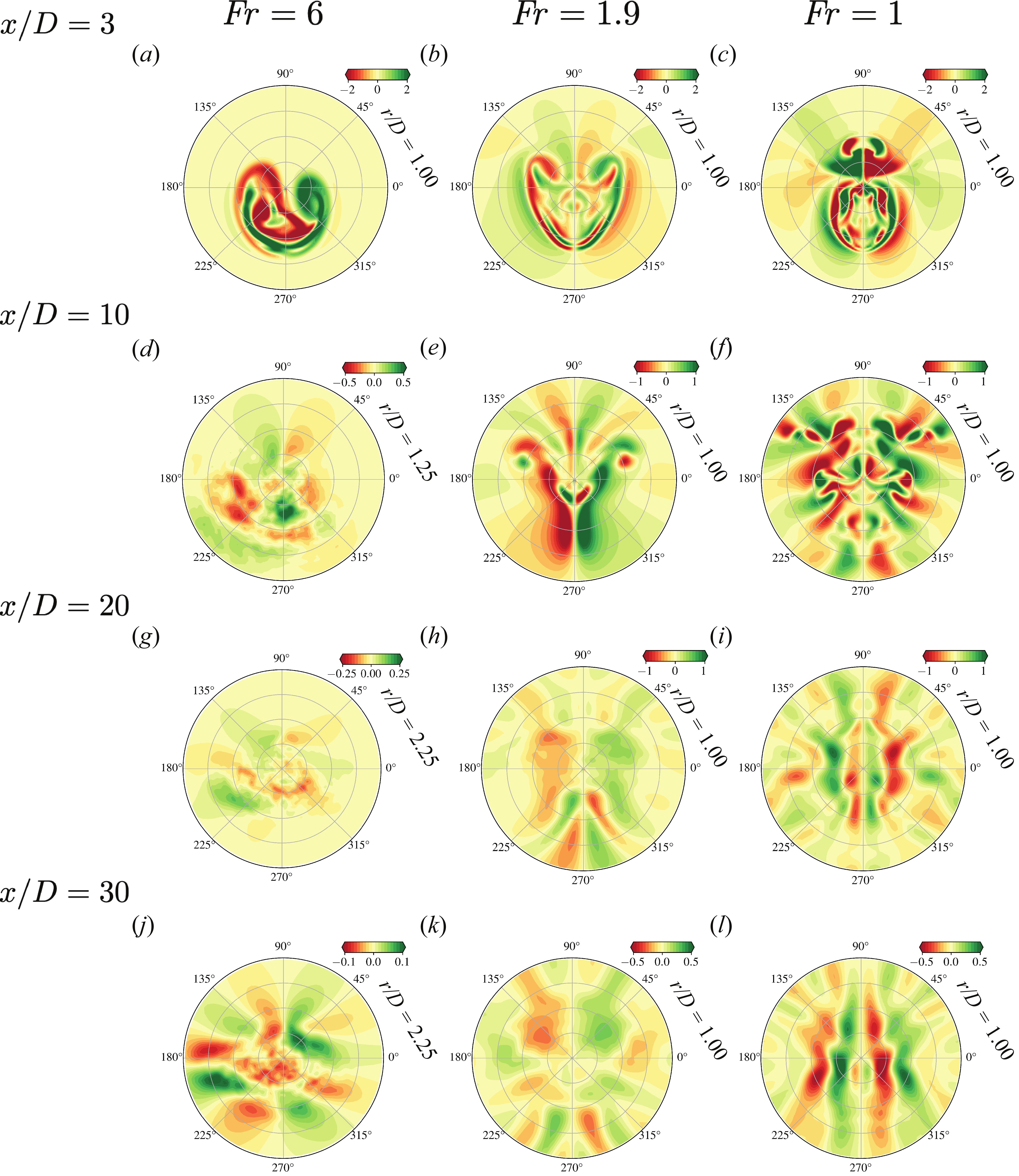}
	\caption{Mean streamwise vorticity $\langle \omega_x \rangle$ at different streamwise locations is compared across three levels of stratification. }
	\label{fig:omgx_mean_strat}	
\end{figure*}

Figure \ref{fig:mean_defect_contours_aoa10} shows that, in response to density stratification,  the wake of the prolate spheroid at an angle of attack evolves in a complex fashion, both in terms of its  lengthscales and the coordinates of its center.  Figure \ref{fig:mean_length_mke_aoa10} presents the evolution of the vertical coordinate of the wake center  and it vertical length scale as a function of $x/D$. Following \cite{brucker_comparative_2010} and \cite{de_stadler_simulation_2012}, these quantities are calculated as follows:
\begin{equation}
	Y^{C} = \frac{\int yU_d^{2}\mathrm{dA}}{\int U_d^{2}\mathrm{dA}}, \qquad Z^{C} = \frac{\int zU_d^{2}\mathrm{dA}}{\int U_d^{2}\mathrm{dA}},
\end{equation}	
\begin{equation}
	L_y^{2} =  \frac{\int (y-Y^{C})^{2}U_d^{2}\mathrm{dA}}{\int U_d^{2}\mathrm{dA}}, \qquad L_z^{2} =  \frac{\int (z-Z^{C})^{2}U_d^{2}\mathrm{dA}}{\int U_d^{2}\mathrm{dA}},
\end{equation}
where $Y^{C}$ and $Z^{C}$ are the wake centers in the spanwise and vertical direction while $L_y$ and $L_z$ are the wake lengthscales in the spanwise and vertical direction.

  Buoyancy strongly inhibits the vertical drift of the wake center, 
shown by Figure \ref{fig:mean_length_mke_aoa10} (a), so that the wake is approximately centered for all cases, including the weaker stratification of $\Fro = 6$.  Although the near-wake development of the wake center at $\Fro = 6$ wake is similar  to the unstratified case, its further development is quite different:  $Z^{C}$  stays very close to the centerline, i.e., $Z^C \approx 0$. There is a lee-wave-induced oscillation with wavelength of $\lambda/D = 2\pi\Fro$  in the evolution of the wake center and its vertical size. At $Fr = 6$, the domain size is not sufficient to show an entire wavelength, Figure \ref{fig:mean_length_mke_aoa10} (a). Interestingly,  $Z^{C}$ of the $\Fro = 1$ wake also stays close to zero, although the mean wake contours show two distinct regions - a primary and a secondary wake - in the left column of \ref{fig:mean_defect_contours_aoa10}(\textit{d,h,l,p}).

 For the present 6:1 spheroid, a pair of counter-rotating vortices with vorticity Reynolds number about $15,000$ is shed in the unstratified case. They maintain their identify as two distinct patches of opposite-signed $\omega_x$ even as their strength decreases with increasing streamwise distance owing to turbulent diffusion and interaction with the wake. 
Figure \ref{fig:mean_omgx_contours_frinf} presents the time-averaged streamwise vorticity $\langle \omega_x \rangle$ contours at four locations in the wake, $x/D = 3, 10, 20$ and $30$. The black circle and cross show the centroid of the positive and negative vortex filaments, given as follows: 
\begin{equation}
Y^{C(+,-)} = \frac{\int y\langle \omega_x \rangle^{(+,-)}\mathrm{dA}}{\int \langle \omega_x \rangle^{(+,-)}\mathrm{dA}}, \qquad Z^{C(+,-)} = \frac{\int z\langle \omega_x \rangle^{(+,-)}\mathrm{dA}}{\int \langle \omega_x \rangle^{(+,-)}\mathrm{dA}},
\end{equation}
where $+$ and $-$ denote the positive and the negative vortex filament respectively.

\begin{figure}
	\centering
	\includegraphics[width=.9\linewidth,keepaspectratio]{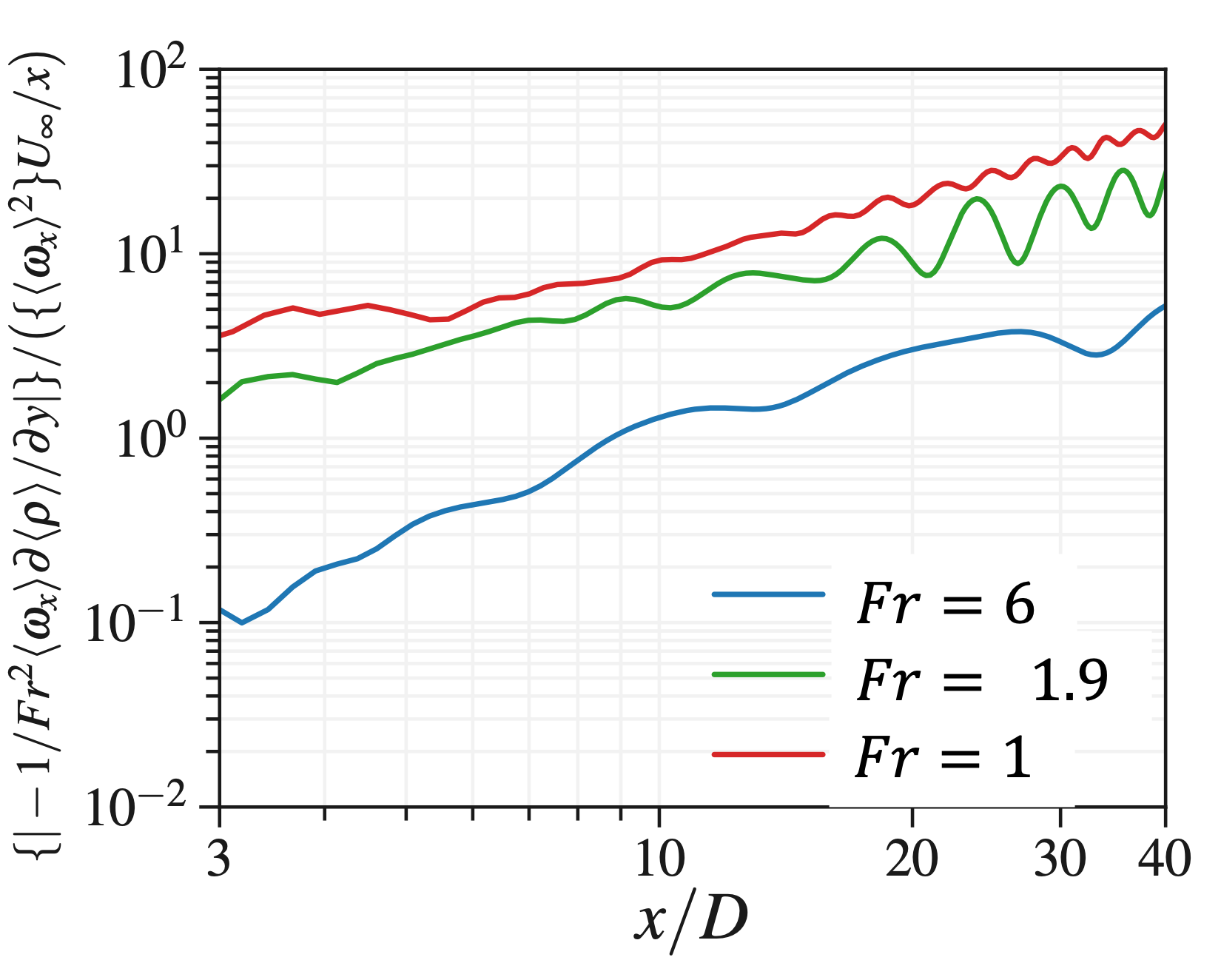}
	\caption{Variation of area-integrated absolute baroclinic term ($|-1/\Fro^2 \langle \omega_x \rangle \partial \langle \rho \rangle/\partial y|$) in the streamwise mean enstrophy equation. This term is normalized by the Lagrangian rate of change of its corresponding mean streamwise enstrophy, $\{\langle\omega_x^{2} \rangle\}U_\infty/x$. $\{ . \}$ denotes the area-integration over a circular cross-section of radius $r/D = 2.$ }
	\label{fig:langrangian_baroclinic_prod}	
\end{figure}

As the vortex pair departs from the body at $x/D = 3$, its distribution exhibits asymmetry, as previously mentioned. The negative and positive filaments are intricately intertwined due to the initial lateral asymmetry. Consequently, we observe lateral shifts in their centroids as well as a downward descent as the pair evolves. It is worth noting that despite the spatial asymmetry in the organization of the negative and positive filaments, the magnitudes of their time-averaged circulation remain equal throughout the computational domain, i.e., $\langle \Gamma^{+} \rangle = -\langle \Gamma^{-} \rangle$. By $x/D = 10$ (figure \ref{fig:mean_omgx_contours_frinf}$b$), these initially intertwined filaments separate into two distinct blobs of negative and positive vorticity, and remain distinct until the end of the computational domain.

Stratification qualitatively alters the topology of streamwise vorticity as shown in figure~\ref{fig:omgx_mean_strat}. At $\Fro = 6$, although there are two distinct vortices at the trailing edge ($x/D = 3$) of the spheroid,  these vortices rapidly disintegrate by $x/D = 10$ in contrast to the unstratified case. Note that the shed vortices at $x/D =3$ have lateral asymmetry that is consistent with the corresponding mean velocity shown in figure~\ref{fig:mean_defect_contours_aoa10} (b). At $\Fro = 1.9$ and 1, flow separation is affected,  the cross flow (with vertical component) is intensified and $\omega_x$ takes the form of two thin counter-rotating strips at $x/D = 3$. Each strip also has an oppositely-signed strip adjacent to it. The $\Fro =  1$ case has a pronounced patch of vorticity above the spheroid corresponding to the secondary wake shown in figure~\ref{fig:mean_defect_contours_aoa10} (c).  Notably, by $x/D = 30$ all stratified cases have $\omega_x$ that is distributed across the wake and in internal gravity waves instead of the two distinct patches in the unstratified case.

%%%
Diagnosis of the vorticity balance  shows that the baroclinic torque  term  plays an important role in  evolution of the streamwise vorticity. Figure \ref{fig:langrangian_baroclinic_prod} shows the variation of cross-section area-integrated absolute baroclinic term in the mean streamwise enstrophy equation, normalized by the Lagrangian rate of change of mean streamwise enstrophy. Normalization by the Lagrangian change allows us to quantify the importance of the baroclinic term to the change of streamwise vorticity. Figure \ref{fig:langrangian_baroclinic_prod} shows that the baroclinic torque has an $O(1)$ effect on the evolution of streamwise enstrophy at $x/D = 8$ and onward in the $\Fro = 6$ wake. Consequently, the the mean vorticity at $\Fro = 6$ (figure~\ref{fig:omgx_mean_strat}, left column) is qualitatively altered relative to the unstratified case by $x/D = 10$ 
(figure \ref{fig:mean_omgx_contours_frinf} b)

\vspace{-5mm}
\section*{ACKNOWLEDGEMENTS}
We gratefully acknowledge the support of the Office of Naval Research grant N0014-20-1-2253.

\vfill\eject
\bibliographystyle{tsfp}
\bibliography{JoseSNH34_references,Sheel_tsfp12_refs,jfm_spheroid_ar6_aoa}
%\bibliography{tsfp}
% In this example, BibTeX is used
% For users not familiar with LaTeX, the bibliography can be typed in directly. In this case, comment the two lines above.
%%%%%%%%%%%%%%%%%%%%%%%%%%%%%%%%%%%%%%%%%%%%%%%%%%%%%%%%%%%%%%%%%%%%%%%
%\section*{SAMPLE REFERENCES}
%
%Kwon, O. K., and Pletcher, R. H., 1981, "Prediction of the Incompressible Flow Over a Rearward-Facing Step", Technical Report HTL-26, CFD-4, Iowa State Univ., Ames, IA.
%
%Lee, Y., Korpela, S. A., and Horne, R. N., 1982, "Structure of Multi-Cellular Natural Convection in a Tall Vertical Annulus," Proceedings, 7th International Heat Transfer Conference, U. Grigul et al., ed., Hemisphere Publishing Corp., Washington, D.C., Vol. 2, pp. 221-226.
%
%Sparrow, E. M., 1980a, "Fluid-to-Fluid Conjugate Heat Transfer for a Vertical Pipe - Internal Forced Convection and External Natural Convection", ASME Journal of Heat Transfer, Vol. 102, pp. 402-407.
%
%Sparrow, E. M., 1980b, "Forced-Convection Heat Transfer in a Duct Having Spanwise-Periodic Rectangular Protuberances", Numerical Heat Transfer, Vol. 3, pp. 149- 167.
%
%Tung, C. Y., 1982, Evaporative Heat Transfer in the Contact Line of a Mixture, Ph.D. Thesis, Rensselaer Polytechnic Institute, Troy, NY.

%%%%%%%%%%%%%%%%%%%%%%%%%%%%%%%%%%%%%%%%%%%%%%%%%%%%%%%%%%%%%%%%%%%%%%

\end{document}